\documentstyle[12pt,aps,prd,preprint]{revtex}
\begin{document}
\title{Gravitation, the Quantum, and Bohr's Correspondence Principle}
\author{Shahar Hod}
\address{The Racah Institute for Physics, The
Hebrew University, Jerusalem 91904, Israel}
\date{\today}
\maketitle

\begin{abstract}
The black hole combines in some sense both the ``hydrogen atom'' and
the ``black-body radiation'' problems of quantum gravity. This analogy
suggests that black-hole quantization may be the key to a quantum
theory of gravity.
During the last twenty-five years evidence has been 
mounting that black-hole surface area is 
indeed {\it quantized}, with  {\it uniformally} spaced area eigenvalues. 
There is, however, no general agreement on the {\it spacing} of 
the levels. 
In this essay we use Bohr's correspondence principle to 
provide this missing link. 
We conclude that the fundamental area unit is $4\hbar\ln3$. 
This is the unique spacing 
consistent both with the area-entropy {\it thermodynamic} relation for 
black holes, with Boltzmann-Einstein formula in 
{\it statistical physics} and with {\it Bohr's correspondence principle}.
\end{abstract}
\bigskip

Everything in our past experience in physics tells us that general
relativity and quantum theory must be approximations, special limits
of a single, universal theory. However, despite the flurry of research, which
dates back to the 1930s, we still lack a complete theory of quantum gravity.
It is believed that black holes may play a major role in our 
attempts to shed some light on the nature of a 
quantum theory of gravity (such as the role 
played by atoms in the early development of quantum mechanics).

The quantization of black holes was proposed long ago in the
pioneering work of Bekenstein \cite{Beken1}. The idea was based on the
remarkable observation that the horizon area of nonextremal black
holes behaves as a classical {\it adiabatic invariant}. In the spirit
of Ehrenfest principle, any classical adiabatic invariant
corresponds to a quantum entity with a {\it discrete} spectrum,
Bekenstein conjectured that the horizon area of a quantum
black hole should have a discrete eigenvalue spectrum.

To elucidate the {\it spacing} of the area levels it is instructive to
use a semiclassical version of Christodoulou's reversible 
processes. 
Christodoulou \cite{Chris} showed that the assimilation 
of a neutral ({\it point}) particle by a (nonextremal) black hole
is reversible if it is injected at the {\it horizon} from 
a radial {\it turning point} of its motion. 
In this case the black-hole
surface area is left unchanged and the changes in the 
other black-hole parameters
(mass, charge, and angular momentum) can be undone by another suitable
(reversible) process. (This result was later 
generalized by Christodoulou and Ruffini 
for charged point particles \cite{ChrisRuff}).

However, in a {\it quantum} theory the particle cannot be
both at the horizon and at a turning point of its motion; 
this contradicts the {\it Heisenberg quantum uncertainty principle}.
As a concession to a quantum theory Bekenstein \cite{Beken2}
ascribes to the particle a {\it finite} effective proper radius
$b$. This implies that the capture process (of a neutral particle) 
involves an unavoidable 
increase $(\Delta A)_{\min}$ in the horizon area \cite{Beken2}:

\begin{equation}\label{Eq1}
(\Delta A)_{\min}=8\pi(\mu^2+P^2)^{1/2}b\  ,
\end{equation}
where $\mu$ and $P$ are the rest mass and 
physical radial momentum (in an orthonormal tetrad) of the particle, respectively.
In the classical case the limit $b\to0$ recovers Christodoulou's 
result $(\Delta A)_{\min}=0$ for a reversible process. 
However, a {\it quantum} particle is subjected to a quantum
uncertainty -- the particle's center of mass 
cannot be placed at the horizon with accuracy better than the 
radial position uncertainty $\hbar/(2\delta P)$.
This yields a lower bound on the increase in
the black-hole surface area due to the assimilation of a (neutral)
test particle

\begin{equation}\label{Eq2}
(\Delta A)_{\min}=4\pi{l_p}^2\  ,
\end{equation}
where $l_p=\left({G\over{c^3}}\right)^{1/2} {\hbar}^{1/2}$ is the
Planck length (we use gravitational units in which $G=c=1$). 
Thus, for nonextremal black holes there is a {\it universal} (i.e., independent of the
black-hole parameters) minimum area increase as soon as one introduces
quantum nuances to the problem.

The universal lower bound Eq. (\ref{Eq2}) derived by Bekenstein is
valid only for {\it neutral} particles \cite{Beken2}. Expression 
(\ref{Eq1}) can be generalized for a {\it charged} particle of rest
mass $\mu$ and charge $e$. Here we obtain

\begin{equation}\label{Eq3}
(\Delta A)_{min}=
\left\{ \begin{array}{l@{\quad,\quad}l}
4\pi [2(\mu^2+P^2)^{1/2} b-e\Xi_+ b^2]&b<b^*\  , \\
4\pi (\mu^2 +P^2)/e\Xi_+&b\geq b^*\  ,
\end{array} \right.
\end{equation}
where $\Xi_+$ is the black-hole electric field (we assume 
that $e\Xi_+>0$) and $b^*\equiv(\mu^2+P^2)^{1/2}/e\Xi_+$.
Evidently, the increase in black-hole surface area 
can be {\it minimized} by maximizing the black-hole electric field. Is
there a physical mechanism which can prevents us from making 
expression (\ref{Eq3}) as small as we wish ? The answer 
is ``yes'' ! {\it Vacuum polarization} effects set an 
upper bound to the strength of the black-hole electric field; the critical
electric field $\Xi_c$ for pair-production of particles with rest mass $\mu$
and charge $e$ is $\Xi_c=\pi\mu^2/e\hbar$ \cite{ScPaDa}.
Therefore, the minimal black-hole area increase is given by 

\begin{equation}\label{Eq4}
(\Delta A)_{\min}=4{l_p}^2\  .
\end{equation}
Remarkably, this lower bound is {\it independent} of the black-hole parameters.

The underling physics 
which excludes a completely reversible 
process (for neutral particles) is the 
{\it Heisenberg quantum uncertainty principle} \cite{Beken2}. 
However, for {\it charged}
particles it must be supplemented by another physical 
mechanism -- a {\it Schwinger discharge} of the black hole 
(vacuum polarization effects). Without this physical process one 
could have reached the reversible limit. It seems that nature has
``conspired'' to prevent this.

It is remarkable that the lower bound found for 
charged particles is 
of the same order of magnitude as the one given by
Bekenstein for neutral particles, even though they 
emerge from {\it different} physical mechanisms.
The {\it universality} of the fundamental lower bound (i.e., 
its independence on the black-hole
parameters) is clearly a strong evidence in favor of 
a {\it uniformly} spaced area spectrum for quantum
black holes. Hence, one concludes that 
the quantization condition of the black-hole surface area 
should be of the form

\begin{equation}\label{Eq5}
A_n=\gamma {l_p}^2\cdot n\ \ \ ;\ \ \ n=1,2,\ldots\ \  ,
\end{equation}
where $\gamma$ is a dimensionless constant.

It should be recognized that the precise values of the 
universal lower bounds Eqs. (\ref{Eq2}) and (\ref{Eq4}) can be
challenged. This is a direct consequence of the {\it inherent fuzziness} of
the uncertainty relation. Nevertheless, it should be clear that 
the fundamental lower bound must be of the same order 
of magnitude as the one given by
Eq. (\ref{Eq4}); i.e., we must have $\gamma=O(4)$.
The small uncertainty in the value of $\gamma$ is the price we must
pay for not giving our problem a full quantum treatment.
In fact, the above analyses 
are analogous to the well known semiclassical derivation 
of a lower bound to the ground state energy of 
the hydrogen atom (calculated by using Heisenberg's 
uncertainty principle, {\it without} solving explicitly the Schr\"odinger
wave equation). The analogy with 
usual quantum physics suggests the next step -- 
a {\it wave} analysis of black-hole perturbations.

The evolution of small perturbations of a black hole 
are governed by a
one-dimensional Schr\"odinger-like wave 
equation (assuming a time dependence of the
form $e^{-iwt}$) \cite{RegWheel}. Furthermore, it was noted that, at late times, all
perturbations are radiated away in a manner reminiscent of the last
pure dying tones of a ringing 
bell \cite{CrViDa}. To describe
these free oscillations of the black hole the notion of quasinormal
modes was introduced \cite{Press}. The quasinormal mode 
frequencies (ringing frequencies) are characteristic of the black hole itself.

It turns out that there exist an 
infinite number of quasinormal
modes for $n=0,1,2,...$ characterizing oscillations with decreasing
relaxation times (increasing imaginary part) \cite{LeBa}. On 
the other hand, the real part of the frequency approaches a {\it constant}
value as $n$ is increased.

Our analysis is based on {\it Bohr's correspondence principle} (1923):
``transition frequencies at large quantum numbers should equal
classical oscillation frequencies''. Hence, we are interested in the
asymptotic behavior (i.e., the $n\to\infty$ limit) of the ringing
frequencies.  These are the highly damped black-hole oscillations 
frequencies, which are compatible with the statement (see, for
example, \cite{BekMuk}) ``quantum transitions do not take time'' 
(let $w=w_R-iw_I$, 
then $\tau\equiv{w_I}^{-1}$ is the effective 
relaxation time for the black hole to return to a quiescent
state. Hence, the relaxation time $\tau$ is 
arbitrarily small as $n\to\infty$).

Nollert \cite{Nollert} found that the asymptotic
behavior of the ringing frequencies of a Schwarzschild black hole is
given by

\begin{equation}\label{Eq6}
Mw_n=0.0437123-{i\over4} \left(n+{1\over2}\right)
+O\left[(n+1)^{-{1/2}}\right]\  .
\end{equation}
It is important to note that the asymptotic limit is {\it independent} of the
multipole index $l$ of the perturbation field. This is a crucial 
feature, which is consistent with the
interpretation of the highly damped ringing 
frequencies (in the $n\gg1$ limit) as being characteristics 
of the {\it black hole} itself.  
The asymptotic behavior Eq. (\ref{Eq6}) was later
verified by Andersson \cite{Andersson} using an 
independent analysis. 

We note that the numerical 
limit $Re(Mw_n)\to0.0437123$ (as $n\to\infty$)
agrees (to the available data given in \cite{Nollert}) with the 
expression $\ln3/(8\pi)$. This identification is supported by
thermodynamic and statistical physics arguments discussed below.
Using the relations $A=16 \pi M^2$ and $dM=E=\hbar w$ one 
finds $\Delta A=4{l_p}^2\ln3$. Thus, we conclude that the
dimensionless constant $\gamma$ appearing 
in Eq. (\ref{Eq5}) is $\gamma=4\ln3$ and the area
spectrum for a quantum black hole is given by
 
\begin{equation}\label{Eq7}
A_n=4 {l_p}^2\ln3\cdot n\ \ \ ;\ \ \ n=1,2,\ldots\ \  . 
\end{equation}

This result is remarkable from a {\it statistical physics} point of
view !  The semiclassical versions of
Christodoulou's reversible processes, which
naturally lead to the conjectured area spectrum
Eq. (\ref{Eq5}), are at the level of mechanics, not statistical
physics. In other words, these arguments did not relay in any way on
the well known thermodynamic relation between black-hole surface area
and entropy.  In the spirit of Boltzmann-Einstein formula in
statistical physics, Mukhanov and Bekenstein \cite{Muk,BekMuk}
relate $g_n\equiv exp[S_{BH}(n)]$ to the number of microstates of the
black hole that correspond to a particular external macrostate
($S_{BH}$ being the black-hole entropy).  Namely, $g_n$ is the
degeneracy of the $n$th area eigenvalue.  The accepted thermodynamic
relation between black-hole surface area and entropy \cite{Beken2} can
be met with the requirement that $g_n$ has to be an integer for every
$n$ only when
 
\begin{equation}\label{Eq8}
\gamma =4\ln k\ \ \ ;\ \ \ k=2,3,\ldots\ \  .
\end{equation}
Thus, statistical physics arguments force 
the dimensionless constant $\gamma$ in Eq. (\ref{Eq5}) to 
be of the form Eq. (\ref{Eq8}). Still, a specific value of $k$
requires further input, which was not aveliable so far. The
correspondence principle provides a first independent derivation of 
the value of $k$.  It should be
mentioned that following the pioneering work of Bekenstein
\cite{Beken1} a number of independent calculations (most of them in
the last few years) have recovered the uniformally spaced area
spectrum Eq.  (\ref{Eq5}) \cite{KMLPLBMK}.
However, there is no general agreement on the spacing of the levels.
Moreover, {\it non} of these calculations is compatible with the
relation $\gamma =4\ln k$, which is a direct consequence of the
accepted thermodynamic relation between black-hole surface area and
entropy. 

The fundamental area spacing $4 {l_p}^2\ln3$ is 
the {\it unique} value consistent both with the area-entropy
{\it thermodynamic} relation, with {\it statistical physics} 
arguments (namely, with the Boltzmann-Einstein formula), 
and with {\it Bohr's correspondence principle}.

\bigskip
\noindent
{\bf ACKNOWLEDGMENTS}
\bigskip

It is a pleasure to thank Jacob D. Bekenstein and Tsvi Piran for
stimulating discussions. 
This research was supported by a grant from the Israel Science Foundation.

\end{document}